\documentstyle[aps,preprint]{revtex}

\begin{document}

\preprint{SNUTP 96-089;~~~
KNUTH-34}

\title{Coherent State Representation of
Semiclassical Quantum Gravity}

\author{Sang Pyo Kim\footnote{Electronic
address: sangkim@knusun1.kunsan.ac.kr}}
\address{Department of Physics\\
Kunsan National University\\
Kunsan 573-701, KOREA}

\maketitle
\begin{abstract}
We elaborate the recently introduced asymptotically
exact semiclassical quantum
gravity derived from the Wheeler-DeWitt equation
by finding a particular coherent state representation
of a quantum scalar field in which the back-reaction of
the scalar field Hamiltonian exactly gives rise to  the classical
one. In this coherent state representation classical
spacetime emerges naturally from semiclassical quantum gravity.
\end{abstract}

\newpage

Canonical quantum gravity based on the
Wheeler-DeWitt (WDW) equation has been used 
as one of its applications to provide
a self-consistent theory of quantum fields in
a curved spacetime (for review and references, see
\cite{kiefer}).
Most of the methods in this direction
rely on the Born-Oppenheimer
idea that the larger mass scale out of 
several mass scales of a quantum system
becomes first semiclassical
and the relatively smaller mass scales
retain the quantum mechanical nature. 
In this approach\footnote{
We distinguish this approach so-called semiclassical
quantum gravity from conventional quantum field theory
in curved spacetimes. We refer to \cite{birrel}
for review and references of the latter approach.}
 to quantum field theory in the curved spacetime
the classical spacetime emerges from the quantum domain but the matter
fields (typically scalar fields)
keep their quantum mechanical properties satisfying the time-dependent
functional Schr\"{o}dinger equation. The advantage of this approach
is that one may treat semiclassical theory relatively simply, 
including some parts of quantum gravitational corrections to the 
matter fields and replacing the energy-momentum tensor by its quantum 
mechanical expectation values. The disadvantage is that the quantum 
corrections of gravity to matter fields may not be achieved fully 
and the renormalization  problem of wave functions may not be 
resolved in this approach to semiclassical quantum gravity.
This approach to semiclassical quantum gravity, despite its shortcomings,
proves quite useful when one considers the quantum gravity
effects semiclassically and especially in the context of cosmology.

However, there has been one step not
completely resolved in deriving
classical gravity from the
WDW equation. One usually assumes that classical gravity 
could be directly obtained from the WDW equation as
the Hamilton-Jacobi equation. It could, of course,
presumably sound physical in a certain domain,
but the logical steps are not fulfilled,
since considering the different mass scales of the gravitational fields 
and matter fields it would be more correct for the
gravity to emerge first
from the quantum domain and then the larger mass scaled matter fields.
The scheme of deriving semiclassical quantum gravity $G_{\mu \nu} = 8 \pi
<\hat{T}_{\mu\nu}>$ from canonical quantum
gravity $ \hat{G}_{\mu \nu} = 8 \pi
\hat{T}_{\mu \nu}$ 
and then classical gravity $G_{\mu\nu} = 8 \pi T_{\mu\nu} $ 
from semiclassical quantum gravity needs a rigorous foundation.

In the previous papers \cite{kim1} we developed an asymptotic
method to derive the quantum field theory of matter fields in 
a curved spacetime from canonical quantum gravity based the
WDW equation. In the case of the quantum
Friedmann-Robertson-Walker (FRW) cosmology minimally coupled to 
a massive scalar field we were able to construct a particular 
Fock space on which the expectation value of the scalar field 
Hamiltonian operator has the same form in terms of  
a complex classical solution as that of classical one \cite{kim2}.

In this paper, we elaborate further that scheme of deriving 
the semiclassical quantum and classical gravity from canonical quantum
gravity based on the WDW equation. It is found that the coherent states
constructed on the particular Fock space \cite{kim2}
make matter fields emerge as classical variables.
In this scheme one is able to show how semiclassical quantum gravity,
the quantum field theory of matter fields, can derive from canonical 
quantum gravity and also how classical gravity theory, the gravity 
coupled to classical matter fields,
can emerge from semiclassical quantum theory of gravity.
For simplicity we shall consider the quantum FRW cosmology minimally
coupled to a massive scalar field. One of the
reasons for specifying the
scalar field is that the coherent states of the massive scalar field 
in a curved spacetime can be constructed
explicitly as will be shown in this paper.

As a simple quantum cosmological model,
we consider a FRW cosmology with the metric
\begin{equation}
ds^2 = - N^2 dt^2 + a^2 d{\sigma}^2_k.
\end{equation}
The action for the FRW cosmology minimally coupled to 
a homogeneous and isotropic massive scalar field is
\begin{equation}
I = \int dt \Biggl[ - \frac{3 m_P^2}{8 \pi}
a^3 \Biggl(\frac{1}{N}
\Bigl( \frac{\dot{a}}{a}\Bigr)^2  - N \frac{k}{a^2} \Biggr)
+ a^3 \Bigl( \frac{\dot{\phi}^2}{2N}
- \frac{Nm^2}{2} \phi^2 \Bigr) \Biggr],
\label{ac}
\end{equation}
where  $ k = 1, 0, -1$ corresponds
to a closed, flat and open universe,
respectively. In the above action we dropped the surface term.
We used the units system such that
$c = \hbar = 1$ and $\frac{1}{G} = m_P^2$.
The conjugate momenta are
\begin{equation}
\pi_a = - \frac{3 m_P^2}{4 \pi} \frac{a \dot{a}}{N},~~
\pi_{\phi} = \frac{a^3 \dot{\phi}}{N}.
\label{mom}
\end{equation}
From the super-Hamiltonian constraint of the ADM formulation
\begin{equation}
{\cal H} = - \frac{2 \pi}{3 m_P^2} \frac{1}{a} \pi_a^2
- \frac{3 m_P^2}{8 \pi} k a
+ \frac{1}{2 a^3} \pi_{\phi}^2 + \frac{a^3 m^2}{2} \phi^2
= 0
\end{equation}
one obtains the WDW equation
\begin{equation}
\Biggl[  \frac{2 \pi}{3 m_P^2} \frac{1}{a}
\frac{\partial^2}{\partial a^2}
- \frac{3 m_P^2}{8 \pi} k a
- \frac{1}{2 a^3} \frac{\partial^2}{\partial \phi^2}
+ \frac{a^3 m^2}{2} \phi^2
\Biggr] \Psi(a, \phi) = 0.
\end{equation}

One of the questions closely related with the correspondence
between quantum and classical theory in general is now to understand
how and when one may recover classical gravity theory
\begin{equation}
\Bigl(\frac{\dot{a}}{a} \Bigr)^2  + \frac{k}{a^2}
= \frac{8 \pi}{3 m_P^2} \Bigl( \frac{\dot{\phi}^2}{2}
+ \frac{m^2}{2} \phi^2 \Bigr),
\label{cl 1}
\end{equation}
which is obtained by varying the action (\ref{ac}) with respect
to the lapse function $N$ and fixing
the temporal gauge $N = 1$.
The other scalar field equation that
constitutes classical gravity is obtained from
the variation of the action with respect to $\phi$
\begin{equation}
\ddot{\phi} + 3 \frac{\dot{a}}{a} \dot{\phi} + m^2 \phi = 0.
\label{os 1}
\end{equation}
One has frequently used the WKB wave function
$\Psi(a, \phi) = F(a, \phi) e^{i S(a, \phi)}$,
where $F$ is a slowly varying function. The rapidly
changing phase factor satisfies the
Hamilton-Jacobi equation
\begin{equation}
 - \frac{2 \pi}{3 m_P^2} \frac{1}{a}
 \Bigl(\frac{\partial S}{\partial a} \Bigr)^2
- \frac{3 m_P^2}{8 \pi} k a
+ \frac{1}{2 a^3} \Bigl(
\frac{\partial S}{\partial \phi} \Bigr)^2 + \frac{a^3 m^2}{2} \phi^2
= 0.
\end{equation}
By identifying $\frac{\partial S}{\partial a} = \pi_a$
and $\frac{\partial S}{\partial \phi} = \pi_\phi$ in (\ref{mom}),
we recover the classical equation (\ref{cl 1}).
But in this approach to classical gravity, there remains
one problem unexplained that
the large mass scale difference between the gravitational field
and matter field in a later stage of cosmological evolution
makes the gravitational field classical but keeps the matter
field quantum mechanical
following the Born-Oppenheimer idea.
This is the main conceptual idea behind semiclassical
quantum gravity theory.

Below we shall develop an alternative to it,
in which classical gravity can be derived
from semiclassical quantum gravity which in turn
can be derived from quantum gravity based on the WDW equation.
It was shown \cite{kim1} that the semiclassical
quantum gravity derived from the WDW equation consists of
the gravitational field equation
\begin{equation}
\Bigl( \frac{\dot{a}}{a} \Bigr)^2  + \frac{k}{a^2}
= \frac{8 \pi}{3 m_P^2} \frac{1}{a^3}
\left< \hat{H}_m \right>,
\end{equation}
and time-dependent Schr\"{o}dinger equation of scalar field
\begin{equation}
i \frac{\partial}{\partial t} \Phi (\phi, t)
= \hat{H}_m \Phi(\phi,t),
\end{equation}
where
\begin{equation}
\hat{H}_m = \frac{1}{2a^3} \hat{\pi}_{\phi}^2
+ \frac{a^3 m^2}{2} \hat{\phi}^2
\end{equation}
is the scalar field Hamiltonian.
As was shown explicitly and fully in \cite{kim1}
the semiclassical quantum gravity equations are asymptotically exact
in the sense of $\frac{1}{m_P^2}
\rightarrow 0$, i.e  $O(\frac{1}{m_P^2})$, provided
that one chooses the quantum states of the scalar field as
the eigenstates of invariant $\hat{I}_m$ satisfying \cite{lewis}
\begin{equation}
\frac{\partial \hat{I}_m}{\partial t}
- i \bigl[ \hat{I}_m , \hat{H}_m \bigr]
= 0.
\end{equation}
The equivalence between different approaches to semiclassical
quantum gravity was recently shown in \cite{bertoni}. We
shall, however, use the approach of \cite{kim1} in which
it is relatively easier to construct
the coherent state representation compared with the others.

Of many invariants it was found that two particular invariants are very
useful and convenient in constructing the Fock space \cite{kim2}:
\begin{eqnarray}
\hat{A}^{\dagger} (t) =  \phi_c (t) \hat{\pi}_{\phi} 
- a^3 (t) \dot{\phi}_c (t) \hat{\phi},
\nonumber\\
\hat{A} (t) =  \phi_c^* (t) \hat{\pi}_{\phi} 
- a^3 (t) \dot{\phi}_c^* (t) \hat{\phi},
\end{eqnarray}
where $\phi_c$ is a complex solution of
(\ref{os 1}) with the boundary conditions\footnote{The sign of the second
condition is corrected from \cite{kim2}}
\begin{eqnarray}
a^3 (t) \Bigl( \phi_c (t) \dot{\phi}_c^* (t) - 
\phi_c^* (t) \dot{\phi}_c (t) \Bigr) = i,
\nonumber\\
{\rm Im} \Biggl( \frac{\dot{\phi}_c (t)}{\phi_c (t)} \Biggr) > 0.
\end{eqnarray}
In fact $\hat{A}^{\dagger} (t)$ acts as
the creation operator and $ \hat{A} (t)$ as
the annihilation operator on the Fock space of 
number states
\begin{equation}
\hat{A}^{\dagger} (t) \hat{A} (t) |n,t> = n |n,t >.
\end{equation}
The exact quantum states are given by
\begin{equation}
\Phi_n (\phi, t) = e^{- i \omega_n (t)} |n, t>
\end{equation}
where 
\begin{equation}
\omega_n (t) =  \int <n,t|\hat{H}_m - i \frac{\partial}{\partial t} |n,t >,
\end{equation}
is a time-dependent phase factor.

We may find the Bogoliubov transformation between two different times
\begin{eqnarray}
\hat{A}^{\dagger} (t) = u (t) \hat{A}^{\dagger} (t_0) + v (t) \hat{A} (t_0),
\nonumber\\
\hat{A} (t) = v^* (t) \hat{A}^{\dagger} (t_0) + u^* (t) \hat{A} (t_0),
\end{eqnarray}
where
\begin{eqnarray}
u(t, t_0) = i a^3 \Bigl( \dot{\phi}_c (t) \phi_c^* (t_0)
- \phi_c (t) \dot{\phi}_c^* (t_0) \Bigr),
\nonumber\\
v(t, t_0) = i a^3 \Bigl( \phi_c (t) \dot{\phi}_c (t_0)
- \dot{\phi}_c (t) \phi_c (t_0) \Bigr).
\end{eqnarray}
The relation
\begin{eqnarray}
|u(t, t_0)|^2 - |v(t,t_0)|^2 = 1
\end{eqnarray}
can be shown by direct substitution.
The above relation can be parameterized as
\begin{eqnarray}
u(t, t_0) = \cosh \nu e^{- i \theta_u},
\nonumber\\
v(t,t_0) = \sinh \nu e^{ - i \theta_v}.
\end{eqnarray}
Then we find a unitary transformation
of the creation operators between two different times
\begin{equation}
\hat{A}^{\dagger} (t) = \hat{S}^{\dagger} (t, t_0)
\hat{A}^{\dagger} (t_0) \hat{S} (t,t_0),
\end{equation}
in terms of the squeeze operator
\begin{equation}
\hat{S} (t,t_0) = \exp \Bigl( i\theta_u \hat{A}^{\dagger} (t_0)
\hat{A} (t_0)\Bigr)
\exp\Bigl( \frac{\nu}{2} e^{-i(\theta_u - \theta_v)}
{\hat{A}^{\dagger 2}} (t_0) - {\rm
h.c.} \Bigr).
\end{equation}
The unitary transformation of the annihilation 
operators can also be found similarly by taking the 
hermitian conjugate of that of the creation operators. 
Note that the squeeze operator is a unitary operator. 
This implies the unitary evolution of the operators on the Fock space. 
This does not always means that the Fock representations 
are unitarily equivalent. In the case of a homogeneous
but not isotropic scalar field in the flat universe
which has an infinite volume, it is known
that the vacua at any two different times are
mutually orthogonal and therefore there 
can be an infinitely many unitarily inequivalent
Fock representations \cite{vitiello,kim2}.

We introduce the coherent states on the
Fock space of the eigenstates of the invariant.
At an arbitrary initial time
we define the coherent state by \cite{glauber}
\begin{equation}
\hat{A} (t_0 ) | \alpha, t_0 > = \alpha |\alpha, t_0 >
\end{equation}
where $\alpha$ is a complex number.
In terms of the creation operator acting
on the vacuum state at that time they read that
\begin{equation}
|\alpha , t_0 > = e^{- \frac{|\alpha|^2}{2}}
\sum_{0}^{\infty} \frac{\alpha^n}{\sqrt{n!}}
|n, t_0 >.
\end{equation}
It can be shown that the coherent states also transform unitarily
\begin{equation}
|\alpha , t > = \hat{S}^{\dagger} (t,t_0) |\alpha, t_0>.
\end{equation}
Thus it follows that
\begin{equation}
\hat{A} (t) |\alpha, t > = \alpha |\alpha, t>.
\end{equation}
The action of the creation operator is
the hermitian conjugate of the annihilation operator
$<\alpha, t| \hat{A}^{\dagger} (t) = \alpha^* <\alpha, t|$.
From the relations
\begin{eqnarray}
\hat{\phi} = (-i) \Bigl(\phi_c \hat{A}^{\dagger} (t)
- \phi^*_c \hat{A} (t) \Bigr),
\nonumber\\
\hat{\pi}_{\phi} = (-i a^3) \Bigl(\dot{\phi}_c \hat{A}^{\dagger} (t) 
- \dot{\phi}^*_c \hat{A} (t) \Bigr),
\end{eqnarray}
we can show that $|\alpha, t>$ really
describes the classical trajectory
\begin{equation}
<\alpha, t| \hat{\phi} | \alpha, t>
= {\varphi}_c (t),
\end{equation}
where
\begin{equation}
{\varphi}_c = \frac{\alpha^* \phi_c - \alpha \phi_c^* }{i},
\end{equation}
is a real classical solution.
The expectation value of the position
operator gives indeed the real classical orbit.
The expectation value of the scalar
field Hamiltonian taken with respect to
the coherent state has a simple form
\begin{equation}
<\alpha, t | \hat{H}_m | \alpha, t>
= \frac{a^3}{2} \bigl( \dot{{\varphi}}_c^2  (t) +
m^2 {\varphi}_c^2 (t) \bigr) +
\frac{a^3}{2} \bigl( \dot{\phi}_c^* (t)
\dot{\phi}_c (t) + m^2 \phi_c^* (t) \phi_c (t) \bigr).
\end{equation}
Note that the last two terms of the expectation value 
come from the quantum fluctuation of vacuum
which are removed by the normal ordering of
the operators
\begin{equation}
<\alpha, t |: \hat{H}_m :| \alpha, t>
= \frac{a^3}{2} \bigl( \dot{{\varphi}}_c^2  (t) +
m^2 {\varphi}_c^2 (t) \bigr).
\end{equation}
Remembering that the
coherent state is a superposition of the eigenstates 
of the particular invariant, we see that the decoupling 
theorem of Lewis and Riesenfeld \cite{lewis} between 
off-diagonal terms still holds and
the semiclassical quantum gravity
in the coherent state representation is asymptotically exact.

In summary we elaborated the previous scheme in which canonical
quantum gravity based on the Wheeler-DeWitt equation
leads to semiclassical quantum and classical gravity \cite{kim1,kim2}.
The coherent state representation is found to make
the expectation value of the quantum energy-momentum tensor
reduce to classical one. Even though we showed the coherent state
representation for the quantum Friedmann-Robertson-Walker cosmology
minimally coupled to a free massive scalar field, we put forth
a {\it conjecture that there may exist the
coherent state representations of
semiclassical quantum gravity for
a generic geometry coupled to fundamental
scalar fields such as scalar fields and fermionic fields}.

The result of this paper may have an
important implication and application to cosmology.
Assuming that the Wheeler-DeWitt equation is valid
just below the Planck scale, we can investigate the condition
under which semiclassical quantum and classical gravity
coupled to inflatons and some other quantum fluctuations necessary for
inflation and reheating emerge and hold true.

\acknowledgments

The author would like to acknowledge
the warm hospitality of ICTP and
Institute for Theoretical Physics of University
of Vienna where this paper was completed.
This work was supported by the Korea Science and 
Engineering Foundation under Contract No. 951-0207-56-2.

\end{document}